%% file: gev_parallel_work.tex
\documentclass[preprint]{sig-alternate-05-2015}      %% LaTeX2e document.  
\pdfoutput=1 
\usepackage{graphicx} %% Preamble.  
\usepackage{color} 
\usepackage{amsmath} 
\usepackage{amssymb,amsfonts,amsmath}
\usepackage{algorithmic} 
\usepackage{url}
\usepackage{letltxmacro}
\usepackage{tikz}
\usepackage{caption}
\usepackage{subcaption}
\usepackage{enumitem}
\usepackage{listings}
\makeatletter
\LetLtxMacro\latexincludegraphics\includegraphics
\@ifpackageloaded{tex4ht}
	{\renewcommand{\includegraphics}[2][]{%
		\immediate\write18{s=\Ginput@path; (IFS='{}'; for i in $s; do convert ${i}#2 ${i}#2.eps; done)}%
		\latexincludegraphics[#1]{#2.eps}}}
	{\renewcommand{\includegraphics}[2][]{\latexincludegraphics[#1]{#2}}}
\makeatother

\newcommand*\circled[1]{\tikz[baseline=(char.base)]{
            \node[fill=none,shape=circle,draw,inner sep=2pt*0.5] (char) {{#1}};}}

\usepackage{amsthm}

\newlength\Origarrayrulewidth

\DeclareGraphicsExtensions{.pdf,.png}

\newif\ifGPblacktext 
\GPblacktexttrue 

\newenvironment{longdescription}
  {\begin{description}[style=unboxed]}
  {\end{description}}

\makeatletter
\def\@isbn{\relax}
\makeatother

\toappear{PDSW-DICS '16 WIP, November 14th 2016, Salt Lake City, UT}

\makeatletter
\def\@copyrightspace{
    \@float{copyrightbox}[b]
    \begin{center}
        \setlength{\unitlength}{0.2pc}
            \begin{picture}(100,6) %Space for copyright notice
                \put(0,-0.95){\crnotice{\@toappear}}
            \end{picture}
        \end{center}
    \end@float}
\makeatother

\begin{document}

%\title{Time taken to write data in parallel on Lustre follows extreme statistics}
\title{A parallel workload has extreme variability}

\numberofauthors{4}
%\author{
%\alignauthor
%Ben Trovato\titlenote{Dr.~Trovato insisted his name be first.}\\
%\affaddr{Institute for Clarity in Documentation}\\
%\affaddr{1932 Wallamaloo Lane}\\
%\affaddr{Wallamaloo, New Zealand}\\

\author{%
\alignauthor Richard Henwood\\
\affaddr{ARM, Inc, 1 Enrico Terrace, 5707 Southwest Parkway, Austin, Texas, USA, 78735}\\
\email{richard.henwood@arm.com}
%\and
\alignauthor
N. W. Watkins\\
\affaddr{Centre for the Analysis of Time Series, London School of Economics, Houghton Street, London, UK}\\
\affaddr{Centre for Fusion, Space and Astrophysics, University of Warwick, UK.}\\
\affaddr{Open University, Milton Keynes, UK.}
%\and
\alignauthor
S. C. Chapman\\
\affaddr{Centre for Fusion, Space and Astrophysics, University of Warwick, UK.}\\
\affaddr{Department of Mathematics and Statistics, University of Tromso, Norway},
\affaddr{Max Planck Institute for the Physics of Complex Systems, Dresden, Germany}\\
%\and
\and
\alignauthor
R. McLay%\\
\affaddr{Texas Advanced Computer Center, 10100 Burnet Rd, Austin, Texas, USA}
%}}
}

%\significancetext{We have shown the same techniques that are used to classify
%extreme events in nature can be used to classify events within parallel
%computing environments. Connecting these two areas of interest invites
%application of analytic techniques (from the complexity science community) to
%the current challenge of understanding variability in computing parallel
%workloads. At a basic level, our discovery distinguishes correct from abnormal
%behaviour and provides a baseline of confidence for correct machine operation.
%Knowledge of correct machine behaviour (beyond crude metrics) has previous been
%empirical -- known to only a few individuals with prolonged exposure to
%parallel environments. Our discovery adds science to this practice and points
%to a future with a deeper appreciation of variability on computer systems at
%scale.}

\maketitle

%\begin{article}

\begin{abstract}

In both high-performance computing (HPC) environments and the public cloud, the
duration of time to retrieve or save your results is simultaneously
unpredictable and important to your over all resource budget. It is generally
accepted (``Google: Taming the Long Latency Tail - When More Machines Equals
Worse Results'', Todd Hoff, highscalability.com 2012) , but without a robust
explanation, that identical parallel tasks do take different durations to
complete -- a phenomena known as variability. This paper advances understanding
of this topic. We carefully choose a model from which system-level complexity
emerges that can be studied directly. We find that a generalized extreme value
(GEV) model for variability naturally emerges.  Using the public cloud, we find
real-world observations have excellent agreement with our model. Since the GEV
distribution is a limit distribution this suggests a universal property of
parallel systems gated by the slowest communication element of some sort.
Hence, this model is applicable to a variety of processing and IO tasks in
parallel environments. These findings have important implications, ranging from
characterizing ideal performance for parallel codes to detecting degraded
behaviour at extreme scales.

%This abstraction allows us to
%apply methods from the field of statistical mechanics and a model for variability. 
%Our model predicts the variability in task durations should be approximated by
%a member of the generalized extreme value (GEV) family of distributions. 
%
%We test our model by selecting and analysing a write to a parallel file system as
%our perfect parallel task. The public cloud is used to construct a experimental
%environment and we observe good agreement between our model prediction and
%observation. 

%The duration of time taken to write data in large-scale compute environments
%with multiple users can vary considerably. This variation comes from a number
%of sources, both systematic and transient. The result is highly complex global
%behaviour. A common attitude towards variation is to treat it as acceptable and
%unavoidable within expert defined bounds.
%
%This paper develops a plausible theoretical model for characterising systematic
%variation on a parallel file system. We show that our model predicts that the
%fluctuations in write durations should be approximated by a member of the
%generalized extreme value (GEV) family of distributions. We test our model by
%fitting experimental results obtained at the Texas Advanced Computer Center at
%The University of Texas at Austin.
%
%The fitted model describes write time durations on a correctly functioning
%parallel file system. A simple application of the model is illustrated to
%provide a precise error estimate for a parallel write.

\end{abstract}


%\abbreviations{GEV, generalized extreme value; HPC, high performance computing}

%The variability in parallel task completion time is critical to performance,
%scheduling, and throughput. Variability is more important as scale and cost
%increases, either in HPC environments or the public cloud. 

\section{Introduction}

Where they exist at all, current models for variability of parallel workloads
on HPC systems implicitly assume I/O variability follows a normal distribution
with the mean and standard deviation the only measure of interest
\cite{evans03,kramer03,skinner05,lofstead10,pusukuri12}. An attempt to fit the
tail of task duration to the log-normal distribution has also been made
\cite{wright09} with limited success. \cite{kim15,haque15} point out that
lowering latency for a given service increases competitiveness of that service.
Their work focuses on reducing the tail latency of a parallel task by reducing
the latency of the individual tasks that makeup the parallel task. Beyond these
studies on parallel workloads, there are an increasing number of phenomena in
computer science and beyond that are best modeled by methods of extreme
statistics
\cite{glynn95,asmussen1998,antonio01,choe98,mink09,coles2001,thomasian04,asmussen2008,andersen2007,bhavsar85,lahyani2012,schroeder07,bramwell09,moloney2010}.

%Queueing Theory has proven extreme statistics for various maxima within a
%variety of queuing systems \cite{glynn95,asmussen1998,antonio01,choe98,mink09}.
%These queues have wide applicability to a variety of system but do not model
%parallel operations where smaller parts make up a complete whole.  

%A general model for variability on homogeneous parallel systems that is
%consistent with observations is unknown. Here we show that the performance
%variability of a common IO pattern on a homogeneous parallel system is best
%modeled by a generalized extreme value (GEV) distribution.  Observations on a
%test system are consistent with our GEV model. Because the GEV distribution is
%a limit distribution this suggests a universal property of systems with a
%slowest communication element of some sort. Hence, this model is applicable to
%a variety of processing and IO tasks in HPC environments and the model is
%sufficiently simple to be understood across physics and computer science
%disciplines. These findings have important implications, ranging from idealized
%performance characteristics for parallel codes to detecting degraded behaviour,
%at extreme scales.

\section{Model}

The modern theory of extreme value distributions can be traced back to the
1920's and two mathematicians: Fisher and Tippett. They considered
\cite{fishertippett1928} extreme values of $n$ samples, each of size $m$ drawn
from the same underlying population. Provided the population values are
independent and identically distributed (i.i.d.), they showed that the
distribution of the extreme values (smallest or largest) drawn from
sufficiently large sub-samples, which in turn are drawn from a larger sample,
tended to one of three possible unique asymptotic forms. For a given underlying
distribution e.g. the exponential, the extremal distribution will be one of the
three, in this case the Gumbel distribution (the others are Fr\'echet, to which
the extremes of power laws are attracted, and the Weibull, also well known in
failure rate modeling for example.) The probability density function of the GEV
with location $\mu$, scale $\sigma$, and shape $\xi$ is:

\begin{equation}
P_{\mu,\sigma,\xi}(x) =
\begin{cases}
\exp \left( - \left( 1+\xi \left(\frac{x-\mu}{\sigma}\right)\right)^{-1/\xi} \right) & \text{if } \xi \neq 0 \\

\exp \left( - e^{(\frac{x-\mu}{\sigma})}\right) & \text{if } \xi = 0
\end{cases}
\label{eqn:gevpdf}
\end{equation}

A detailed description, and physical examples of extreme value theory are
presented in \cite{leadbetter1983,coles2001,sornette04}. Next, we choose a
common an simple parallel task (a write to a parallel file system) and argue
that the i.i.d. assumption needed for GEV behavior are directly applicable as
follows:

\begin{longdescription}

\item[The storage nodes are independent.] A storage node is here defined as a
device that receives a portion of a file during a parallel write. While it is
common to collect multiple devices into a storage array, our model treats an
array as a single storage node that is independent from other arrays.

\item[A write task takes place from a single node to many storage nodes.] Of
the many I/O scenarios enumerated in the article \cite{newman08}, this paper is
concerned with the duration to complete \emph{scenario 5: Checkpoint/restart
with large I/O requests}. This is also known as a `one-to-many' operation.

\item[The dominant source of variation within the system arises from the
storage nodes.] The non-dominant sources of latency in the system including:
network switches, network cards, interrupts, kernel buffers, PCI interfaces,
OS schedulers, memory latency etc are all assumed to be comparatively
small. 

\item[The client node is connected to each of the storage nodes by an
identical network connection.] The network connections connecting the client and
storage nodes are identical in bandwidth and latency.

\end{longdescription}

%The experiment (described in detail in Section IV) is designed to embody the
%conditions in which these assumptions will hold true. Given this, the model for
%the duration of time taken to complete a write $T_g$ on a parallel file system
%is constructed as follows:
%
%\spnewtheorem{thm}[theorem]{Postulate}{\bfseries}{\itshape}
%\begin{postulate}
%\begin{thm} 

\section{Experiment}

A quantity of interest to many in HPC is the duration of
time to complete a given task. Our chosen task is a write operation on a
parallel file system with a duration of $T_g$. We assume that there is a
baseline characteristic of the parallel task duration that is observable on a
quiescent system without congestion $T_s$. Congestion is a important factor in
network operations \cite{evans03,thomasian04,calzarossa04} that arises with a
shared network or the storage nodes that are busy with other tasks. We encode
the congestion penalty (which we call background traffic factor) as a constant
of proportionality $k_t$. This gives: $T_g = k_t T_s$. A completely quiet
system without congestion or background traffic is the state where $k_t=1$. If
background traffic is present, $k_t>1$.

We extend our model with the assumptions: an observed file transfer to a
\emph{single} storage node will take $S$ seconds where $S$ is an observation of
the storage node that behaves with a given probability distribution: $p(s)$.
Hence the time taken $T_s$ for the storage nodes to complete a \emph{parallel}
write in our model is the largest value of $S$ from $m$ storage nodes: $T_s =
\max\{S_1, S_2 ... S_m\}$.  By substitution, we arrive at:

\begin{equation}
T_g = k_t\max\{S_1, S_2 ... S_m\}.
\label{eqn:writemodel}
\end{equation}

i.e. a client will observe a write time onto a parallel file system that is
limited by the last storage node to complete the task: $T_g =
P_{\mu,\sigma,\xi}(x)$ from equation \eqref{eqn:gevpdf}.

From Extreme Value Theory, provided  $m$ is sufficiently large and with our
additional constant traffic constraint ($k_t$ is constant across observations),
we construct the following testable hypothesis: the times taken to transfer a
file onto a large number of storage nodes will have a distribution approximated
a random variable that has a extreme value distribution, given a fixed level of
background traffic (congestion) and our previously stated assumptions of the
system hold true.

An investigation to explore the distribution $T_g$ was initially conducted at
TACC on the Ranger system. Encouraging results were obtained. However, these 
results were identified as unreliable
because the experimental run used {\tt dd} with a block size of more
than 2GB. For some configurations (apparently including Ranger), {\tt dd}
will stop writing after 2GB and return success. This initial data was
discarded. An experimental run was subsequently completed on both
Stampeed and Lonestar4 without success: these machines did not include the i.i.d. 
assumptions previously stated.

A second experiment was designed and conducted on the Amazon
Web Services (AWS) public cloud. Cloud based computing has grown in popularity
as a inexpensive tool for research, and performance evaluations are an area of
active research \cite{yao13,bautista12,iosup11,wang10}. AWS allows dynamic
construction of arbitrary configurations as well as isolated network
environments - necessary to ensure constant $k_t$ in our model. For a
completely isolated network with a single client running a single job, $k_t$ =
1.

Amazon Web Services provide basic specifications of the network and storage
performance. They state a throughput of 128 MBps per volume
\footnote{\protect\url{http://aws.amazon.com/ebs/details/}}, 62.5MBps per
instance for write
\footnote{\protect\url{http://docs.aws.amazon.com/AWSEC2/latest/UserGuide/ebs-ec2-config.html}}.
The dynamically constructed cluster was created within a `placement group'
\footnote{\protect\url{http://docs.aws.amazon.com/AWSEC2/latest/UserGuide/placement-groups.html}}.
This is a logical group of instances that enables applications to participate
in a low-latency, 10Gbps network.
Published values for the throughput of c3.large storage servers could not be
obtained. The maximum theoretical bandwidth of a 10Gbps network is 1250 MBps.
The mean value observed in our experiment is 45MBps. From these calculations it
would appear that the instance throughput (possibly on the client) is the
bottleneck in our system configuration.

Our experiments are performed on the Lustre\footnote{Other names and brands may
be claimed as the property of others.} parallel file system version 1.8.9-wc1.
While more recent Lustre software releases are available, using synchronous
write in our experiment prohibited versions of Lustre that do not have a fix
for LU-1669. At the time (Autumn 2015), 1.8 was the most popular Lustre
version that supported parallel direct write. In addition, previous variability
papers have chosen 1.8 for their studies. To avoid complications with caches,
only synchronous write operations are considered in this study. The design of
the Lustre file system version 1.8 requires a serialized meta-data request to
open and close the file. We use a simple code (provided in the appendix) that
measures the time for serialized meta-data requests separately to the parallel
data transfer request. Our experiment defines a single write as a total file
size of 512 MB written to 16 storage nodes. The default stripe size of 1MB was
used. Choosing a files size of 512 MB ensures the file is small enough to fit
in the client memory (total of 7.5GB) without needing costly swapping. 16
storage nodes is chosen as a sufficiently large population ($m$) and a total of
400 observations made to ensure sufficient fidelity of the underlying
distribution and increase confidence of correct identification \cite{henwood08}

Specific compute instance (EC2) types and Elastic Block Store (EBS) were chosen
as shown in Figure 1. %\ref{fig:aws_setup}. 
The cluster was constructed behind a head
node (not shown) in a private subnet within a placement group. The EC2
instances were shared tenancy. All instances in the experimental setup were
CentOS 5.11 with Lustre 1.8.9-wc.

\begin{figure}
\centering \def\svgwidth{6.5cm} % sets the image width, this is optional
%\resizebox {.6\columnwidth} {!} {
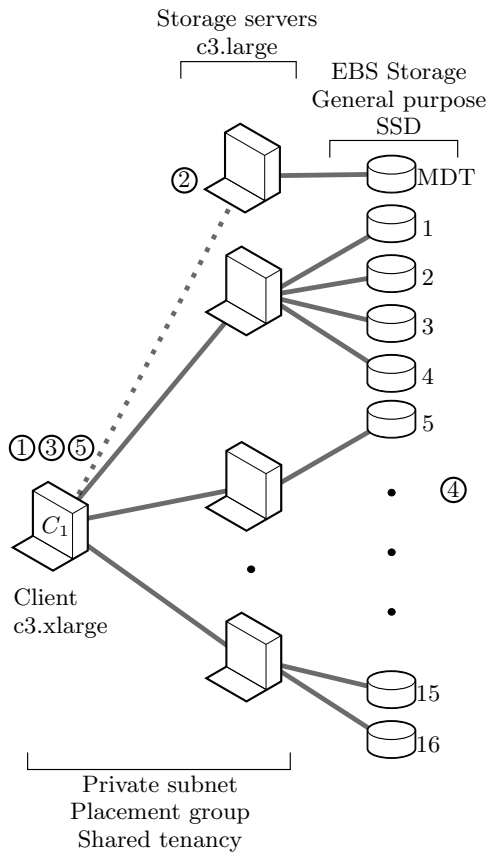
%}
%
\caption{A typical high performance storage architecture with a single client
node $C_1$. Storage targets (1-16) are attached in groups of four to storage
servers. A read or write operation from $C_1$ occurs across all storage targets
in parallel. A write operation includes the following high level steps:
\protect\circled{1} $C_1$ executes a single task and accumulates results in
memory until the task is complete.
\protect\circled{2} $C_1$ requests a file handle from the metadata server. The
metadata server persists data on storage (labelled `MDT') and instructs the
client to write to all the storage nodes during writing. From this point 
onwards the system storage targets behave with i.i.d. characteristics.
\protect\circled{3} A timer begins on $C_1$. $C_1$ and the contents of the
memory is written to all the storage nodes as a synchronous write.
\protect\circled{4} The storage servers pass the data directly through to the
EBS storage nodes (1-16).
\protect\circled{5} The timer is stopped when $C_1$ is told that the write is
complete. The value of the timer is $T_g$.
} 
\label{fig:aws_setup} 
\end{figure}

\section{Results}

Figure 2 %\ref{fig:400obsfits} 
shows the duration of a parallel write is best
approximated by equation \eqref{eqn:writemodel}. This results supports the hypotheses
that  the duration of a parallel write is controlled by the slowest node. GEV
distributions are defined by three parameters: location, scale, and shape. The
result of our work indicates that all three are valuable in capturing the
variability characteristics of a system. HPC performance variability data first
published in \cite{kramer03,skinner05} may now be better explained using the
GEV model. \cite{schwarzkopf12} (and references therein) highlight the under
appreciated importance, and poor level of understanding of variability, within
cloud computing environments. Our results present a model that will provide for
a deeper understanding of variability on both the cloud and HPC.

\begin{figure}
\begin{center}
\resizebox {.75\columnwidth} {!} {
\input{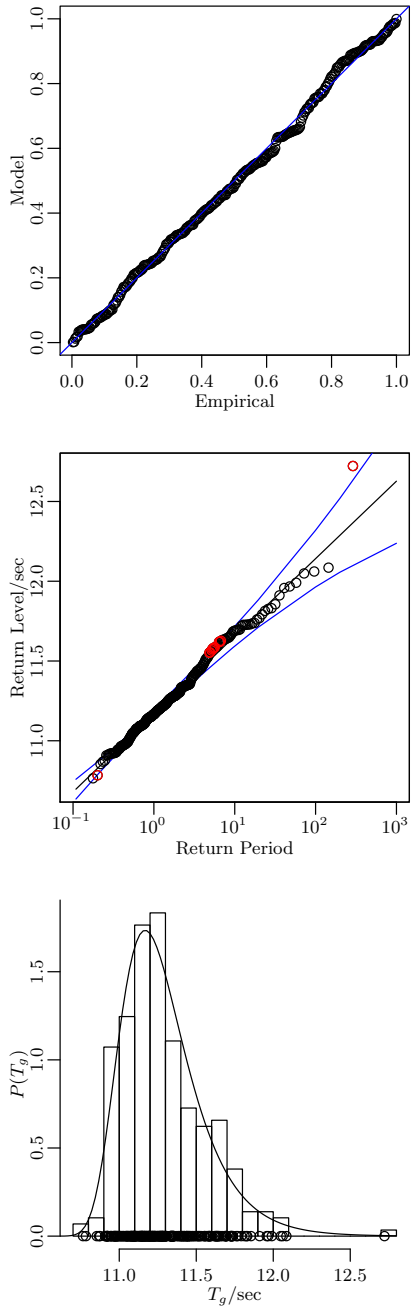}
}
\caption{
Parallel write times follow extreme statistics.  
400 consecutive observations of $T_g$ were taken. The top panel shows the cumulative
value of the observation against the model value. The middle panel is the
observed quantity plotted against the modeled quantity with the 95\% confidence
interval of the value of $\xi$ shown as a blue line. Observations that fall
outside the 95\% confidence interval are colored in red. The bottom panel
presents the observation histogram in 20 equal width bins with the fitted
probability density over-plotted. The GEV fit has location $\mu = 11.1679 \pm
0.0140$, scale $\sigma = 0.2120 \pm 0.0101$, and shape $\xi = -0.00105 \pm
0.0415$. Values of $\mu$, $\sigma$, $\xi$, standard errors, and outliers were
calculated using the {\tt{ismev}} library \cite{ismev} within the R language
environment \cite{rmanual}.}
\label{fig:400obsfits}
\end{center}
\end{figure}

\section{Conclusions}

From extreme value theory, as the number of nodes increases we anticipate a
universal behavior will emerge in systems of this type. We can confirm that
with the conditions already stated, this is the case in our system (Figure 2
%\ref{fig:400obsfits}
). Our idealized experiment has wider implications as it maps onto a large
class of systems, both physical and societal, where the essential element is
waiting for a response in parallel from any nodes. In the computing field, for
example, the Monte Carlo method is widely used and deployed at parallel scale
and under certain configurations, the time to result would be expected to have
a GEV distribution.

A complete, efficient, and accurate model of an HPC system is critical in
optimizing utilization of this limited resource. Queues have already
successful modelling a number of components of an HPC system including task
scheduling \cite{thomasian04,terekhov13}, network systems
\cite{lahyani2012}, and failure and recovery
\cite{asmussen2008}. Our GEV model for parallel
transfer grows the tools available to a model an entire, active, HPC cluster.

The specter of traffic or network congestion is often introduced when looking
at variability in benchmark measurements. If we are benchmarking a parallel
task, and the GEV model is accurate, we expect a tail in the variability even
in the complete absence of traffic. After the underlying variability of a
parallel workload characterized, the affect of network congestion on the same
workload can now be quantified.

As high performance computing continues to develop and increase parallelism,
new libraries become available (and necessary), to simplify interfacing with
data objects \cite{barton13}. For example, the {\tt{t3pio}} library provides
automatic configuration for MPI applications that use HDF5. With the GEV model,
a library can be calibrated for ideal parallel (GEV) behavior and measure
deviations from this behavior as values that are unlikely. The journey to
exascale computing means vast increases node count and parallelism
\cite{agerwala10}. We expect GEV to be a powerful tool in understanding and
exploiting variability on HPC systems in the future.

In summary, this paper explains the variability in parallel writes. The
variability is explained by extreme value theory. Our analysis of data
collected from a parallel write task performed in the public cloud found good
agreement with well understood extreme statistics. Studies of parallel tasks
should perhaps begin to consider examining repeated runs for evidence of
extreme value distribution as a unique parallel performance signature.

%Queueing theory is a perfect fit for modeling a number of
%parts of an HPC system including job schedulers \cite{terekhov13} and network
%congestion \cite{mink09}. By combining our model for parallel I/O with open
%queueing network theory more accurate simulations and higher utilization of
%large scale computer systems may become possible.
%
%The journey to exascale computing means a vase increase in parallelism for
%compute \cite{agerwala10} and storage \cite{barton13} accompanied by
%corresponding growth for error and failure rates. New libraries are developed
%to simplify interfacing with data at massive scale (for example the
%{\tt{t3pio}} library provides automatic configuration to MPI applications that
%use HDF5 on Lustre). With the GEV model, libraries involved with paralle I/O
%can be calibrated for `ideal' variability and automatically identify degraded
%performance more quickly and accurately than non-GEV models.

%As high performance computing continues to develop and increase parallelism,
%new libraries become available, and necessary, to simplify interfacing with
%data objects \cite{barton13}. For example, the {\tt{t3pio}} library provides
%automatic configuration for MPI applications that use HDF5. With the GEV model,
%a library can be calibrated for ideal parallel (GEV) behavior and measure
%deviations from this behavior as values that are unlikely. The journey to
%exascale computing means vast increases node count and parallelism
%\cite{agerwala10}. We expect GEV to be a powerful tool in understanding and
%exploiting variability on HPC systems in the future.

\section{Acknowledgments} 
RH is grateful to Intel High Performance Data
Division who supported this work. RH thanks various anonymous
reviewers who significantly improved this manuscript. SCC is supported by the
UK EPSRC and STFC.

\bibliographystyle{abbrv}
\bibliography{refer}
%\blfootnote{* Other names and brands may be claimed as the property of others}
%\end{article}

%\clearpage
\appendix[Supporting Information: data write code]
\begin{lstlisting}[language=C,float=*,lastline=58]
/* Code to observe GEV variability. Typically, 100 runs should
 * generate sufficient data for GEV to be confidently observed.
 * Example values:
 *     $FILESIZE = 512 // in MiB
 *     $STRIPECOUNT = 16
 *     $TARGETDIR = /mnt/lustre // on a Lustre file system.
 *
 * Ensure striping is set
 *
 *     lfs setstripe -c ${STRIPECOUNT} -s 1M $TARGETDIR"
 *
 * Run with:
 *
 *     ./timed_write $FILESIZE $TOPDIR/$runNum.dat
 *
 * Values of 'write' result, are expected to have GEV distribution.
 */
#include <stdio.h>
#include <stdlib.h>
#include <time.h>
#include <sys/time.h>
#include <fcntl.h>
#include <errno.h>
#include <string.h>

int main(int argc, char *argv[])
{
	int fd;
	int rc;
	unsigned char *bytes;
	long long write_size_mb;
	struct timeval start, end;
	long long calloc_elapsed_l, free_elapsed_l;
	long long open_elapsed_l, write_elapsed_l, close_elapsed_l;
	write_size_mb = atoll(argv[1]) * 1024 * 1024;

	printf("allocating local memory for write of %llu bytes\n", write_size_mb);

	/* gettimeofday has some known limitations:
	 * http://stackoverflow.com/questions/88/
	 * However, it should be sufficiently reliable for this experiment. */
	gettimeofday(&end, NULL);
	bytes  = calloc(write_size_mb, sizeof(unsigned char));
	if (bytes == NULL) {
		printf("can't allocate %s MiB: '%s'\n", argv[1], strerror(errno));
		return 1;
	}

	gettimeofday(&start, NULL);
	calloc_elapsed_l = (start.tv_sec * 1000000 + start.tv_usec) - 
		(end.tv_sec*1000000 + end.tv_usec);
	
	gettimeofday(&start, NULL);
	fd = open(argv[2], O_SYNC|O_WRONLY|O_CREAT, 0644);
	if (fd < 0) {
		printf("can't open file: '%s' error: '%s'\n", argv[2], strerror(errno));
		return 1;
	}
\end{lstlisting}
\begin{lstlisting}[language=C,float=*,firstline=59]
	gettimeofday(&end, NULL);
	open_elapsed_l = (end.tv_sec * 1000000 + end.tv_usec) - 
		(start.tv_sec*1000000 + start.tv_usec);

	rc = write(fd, bytes, write_size_mb * sizeof(unsigned char));
	if (rc < 0) {
		printf("can't write to this file: '%s' error: '%s'\n", argv[2], strerror(errno));
		return 1;
	}
	gettimeofday(&start, NULL);
	write_elapsed_l = (start.tv_sec * 1000000 + start.tv_usec) - 
		(end.tv_sec*1000000 + end.tv_usec);

	close(fd);
	if (fd < 0) {
		printf("can't close file: '%s' error: %s\n", argv[2], strerror(errno));
		return 1;
	}
	gettimeofday(&end, NULL);
	close_elapsed_l = (end.tv_sec * 1000000 + end.tv_usec) - 
		(start.tv_sec*1000000 + start.tv_usec);

	free(bytes);
	gettimeofday(&start, NULL);
	free_elapsed_l = (start.tv_sec * 1000000 + start.tv_usec) - 
		(end.tv_sec*1000000 + end.tv_usec);

    printf("write complete: calloc %lf open %lf write %lf close %lf free %lf total %lf\n", 
		calloc_elapsed_l/1000000.0,
		open_elapsed_l/1000000.0,
		write_elapsed_l/1000000.0,
		close_elapsed_l/1000000.0,
		free_elapsed_l/1000000.0,
		(calloc_elapsed_l
			+ open_elapsed_l 
			+ write_elapsed_l 
			+ close_elapsed_l
			+ free_elapsed_l)/1000000.0);

	return (0);
}
\end{lstlisting}

%\begin{figure*}
%\begin{minipage}[t]{1.0\textwidth}
%\appendix[Supporting Information: Simple data write code]
%\lstinputlisting[language=C]{timed_write.c}
%\end{minipage}
%\end{figure*}

\end{document}

%% file: 1-16_model.pdf_tex
%% Creator: Inkscape inkscape 0.48.5, www.inkscape.org
%% Accompanies image file '1-16_model.pdf' (pdf, eps, ps)
\begingroup%
  \makeatletter%
  \providecommand\color[2][]{%
    \errmessage{(Inkscape) Color is used for the text in Inkscape, but the package 'color.sty' is not loaded}%
    \renewcommand\color[2][]{}%
  }%
  \providecommand\transparent[1]{%
    \errmessage{(Inkscape) Transparency is used (non-zero) for the text in Inkscape, but the package 'transparent.sty' is not loaded}%
    \renewcommand\transparent[1]{}%
  }%
  \providecommand\rotatebox[2]{#2}%
  \ifx\svgwidth\undefined%
    \setlength{\unitlength}{812.78994141bp}%
    \ifx\svgscale\undefined%
      \relax%
    \else%
      \setlength{\unitlength}{\unitlength * \real{\svgscale}}%
    \fi%
  \else%
    \setlength{\unitlength}{\svgwidth}%
  \fi%
  \global\let\svgwidth\undefined%
  \global\let\svgscale\undefined%
  \makeatother%
  \begin{picture}(1,1.72728049)%
    \put(0,0){\includegraphics[width=\unitlength]{1-16_model.pdf}}%
    \put(0.8354035,1.40651615){\color[rgb]{0,0,0}\makebox(0,0)[lt]{\begin{minipage}{0.1108034\unitlength}\centering \footnotesize{MDT}\end{minipage}}}%
    \put(0.00991393,0.54823148){\color[rgb]{0,0,0}\makebox(0,0)[lt]{\begin{minipage}{0.28692697\unitlength}\raggedright Client\\ c3.xlarge\end{minipage}}}%
    \put(0.59057912,1.62423283){\color[rgb]{0,0,0}\makebox(0,0)[lt]{\begin{minipage}{0.41603396\unitlength}\centering EBS Storage General purpose SSD\end{minipage}}}%
    \put(0.03448958,0.69471138){\color[rgb]{0,0,0}\makebox(0,0)[lt]{\begin{minipage}{0.12434823\unitlength}\centering $C_1$\end{minipage}}}%
    \put(0.82505737,0.24426064){\color[rgb]{0,0,0}\makebox(0,0)[lt]{\begin{minipage}{0.06618648\unitlength}\centering $16$\end{minipage}}}%
    \put(0.82485276,1.3020558){\color[rgb]{0,0,0}\makebox(0,0)[lt]{\begin{minipage}{0.06618648\unitlength}\centering $1$\end{minipage}}}%
    \put(0.82485276,1.2019059){\color[rgb]{0,0,0}\makebox(0,0)[lt]{\begin{minipage}{0.06618648\unitlength}\centering $2$\end{minipage}}}%
    \put(0.82485276,1.09980894){\color[rgb]{0,0,0}\makebox(0,0)[lt]{\begin{minipage}{0.06618648\unitlength}\centering $3$\end{minipage}}}%
    \put(0.82485276,0.99771199){\color[rgb]{0,0,0}\makebox(0,0)[lt]{\begin{minipage}{0.06618648\unitlength}\centering $4$\end{minipage}}}%
    \put(0.82505738,0.90191654){\color[rgb]{0,0,0}\makebox(0,0)[lt]{\begin{minipage}{0.06618648\unitlength}\centering $5$\end{minipage}}}%
    \put(0.06665998,0.16626264){\color[rgb]{0,0,0}\makebox(0,0)[lt]{\begin{minipage}{0.48690699\unitlength}\centering Private subnet\\ Placement group\\ Shared tenancy\end{minipage}}}%
    \put(0.21840738,1.73270977){\color[rgb]{0,0,0}\makebox(0,0)[lt]{\begin{minipage}{0.49834065\unitlength}\centering Storage servers\\ c3.large\end{minipage}}}%
    \put(0.02727862,0.82224372){\color[rgb]{0,0,0}\makebox(0,0)[b]{\smash{1}}}%
    \put(0.35951239,1.36989758){\color[rgb]{0,0,0}\makebox(0,0)[b]{\smash{2}}}%
    \put(0.08768476,0.82224372){\color[rgb]{0,0,0}\makebox(0,0)[b]{\smash{3}}}%
    \put(0.90974867,0.7362175){\color[rgb]{0,0,0}\makebox(0,0)[b]{\smash{4}}}%
    \put(0.14809089,0.82224372){\color[rgb]{0,0,0}\makebox(0,0)[b]{\smash{5}}}%
    \put(0.82505737,0.34859264){\color[rgb]{0,0,0}\makebox(0,0)[lt]{\begin{minipage}{0.06618648\unitlength}\centering $15$\end{minipage}}}%
  \end{picture}%
\endgroup%